# Is Facebook regionally a small world network?

Giovanna Maria Dimitri[1,2][0000-0002-2728-4272]

[1] DIISM, Università degli Studi di Siena, Italy

[2] Computer Laboratory, Department of Computer Science, University of Cambridge, UK
giovanna.dimitri@unisi.it

gmd43@cl.cam.ac.uk

**Abstract.** The analysis of social networks, can lead to important discoveries concerning society and trends. Can in fact imply the discovery of several new aspects of social behavior, as well as understanding the interest behind certain topics. Facebook, is now used worldwide, by approximately 3 billions of users, and has become one of the main sources of information. For analyzing facebook data, not only worldwide scales are important, but it is crucial to interpret local data behavior. In this paper we address and analyse Facebook at a regional dimensionality of Facebook, and evaluate the properties of "regional Facebook" as a small world network. To do this, we propose a novel approach a novel experimental setup, to simulate worldwide known Milgram six degrees of separation experiment. The novel repeated sampling random procedure proposed to reproduce the Milgram experiment showed promising and interesting results, allowing to evaluate the regional Facebook scale.

## 1. Introduction

Social networks have deeply shaped our daily lives, our relationships, and our connections to friends and families [8-11]. Complex systems and network analysis [12-14,18-21] have proved to be a very powerful tool to discover and analyze network behaviors in many different field. In social networks, in particular, social experiments and network theory have created a strong tie, to understand human behaviours and trends.

In the study of human relationships, and connections one of the most well known social experiment, was the so called *"six degrees of separation experiment"*. At the end of 1960s, Stanley Milgram, an American psychologist, showed experimentally how two random sampled people in the world are divided by only 5 connections. This is also better known as the 6 degrees of separation or small world theory [6].

Subsequently developed theories of small world networks, have proved to be very important in different contexts. In [15], for example ,the authors analysed the properties of the small world networks and the implications of the spreading of the epidemic in the university campus. In [16] instead the effect of small world networks was analysed towards the understanding of spreading the information and diffusion of innovations. Similarly in [17] the authors present application of small networks properties to the case of the academic connections.

In this paper we propose a novel experimental procedure, based on a repeated sampling procedure, to verify and repeat experimentally the Milgram Experiment in a social and interaction Facebook regional network.

The paper is organized as follows. In section 2 we briefly describe the structure of the dataset used in our experiments. Subsequently in Section 4 we present the experimental analysis on the network and we discuss implications of the results obtained.

Eventually in Section 5 we present the conclusions, considering social and theoretical aspects related to the findings proposed. Future works are briefly sketched and presented.

## 2. The dataset

The dataset used in the present paper describes the social and interaction network of a small regional network of Facebook. This was obtained before 2009 [1], as such datasets are no longer available. After 2009 Facebook, in fact, removed regional features, that is the possibility for people within the same regional network, to have full access to each other's' personal data [2]. The dataset we used in our experiments is composed by two graphs: a social graph and an interaction graph and was retrieved as released and described in publication [2]. We will describe the two separately in the following subsections.

1. **Social Graph**

The social graph is inherently an undirected graph. The graph models the friendship relationship in the Facebook regional subnetwork considered. From Table 1, we can immediately see how the graph presents a quite high number of nodes and edges. We therefore estimated some summary statistics of the graphs, such as the average degree, average node degree, average clustering coefficient and assortativity measure. To gain an understanding of the clustering coefficient of the social graph, we built the corresponding Erdos Renyi random graph [5], with the same number of nodes and probability of an edge described as:

where   is the average node degree and   is the number of nodes. The clustering coefficient of the corresponding Erdos-Renyi graph is 0.000000633539157535. This significantly lower value shows the fact that the Social Facebook Graph analysis is likely to be identified as a Small World Network [5].

*Table 1: Social Interaction Graph Metrics*

| # Nodes | 657681 |
|---|---|
| # Edges | 1302764 |
| **Average Degree** | 1.980 |
| **Average Node Degree** | 3 |
| **Average Clustering Coefficient** | 0.0660255640547 |
| **Assortativity** | -0.299485679564 |

Furthermore in Table 1 we can notice the presence of a negative value for the assortativity measure. Such indicator, is defined as the Pearson correlation coefficient of the degrees of all node pairs, for every edge in the graph. The fact that in this case assumes negative values indicates that nodes tend to connect to others with dissimilar degrees. Considering the nature of our network, this is somehow unexpected, as compared to what generally happens in Social Networks. In fact, in the Social Network cases, usually assortativity tends to assume a positive value, since nodes, with similar degrees
tend to connect to each other [3][4]. A possible explanation of such a peculiar behavior is the *"non-conventional"* structure of the social interaction network. The graph, in fact, is not connected, and composed by a set of 6 connected components of varying dimensions as shown in Table 2.

*Table 2: Social Interaction Graph Connected Components and dimensions*

| COMPONENT | Number of Nodes N |
|---|---|
| 1 | 657587 |
| 2 | 68 |
| 3 | 11 |
| 4 | 7 |
| 5 | 5 |
| 6 | 3 |
| | |

From Table 2, in fact, we can notice the presence of a very big component, that includes the highest percentage of network nodes. As a further statistical visualization plotting the network degree distribution, we can appreciate the presence of a scale free behavior, as it is shown in Figure 1.

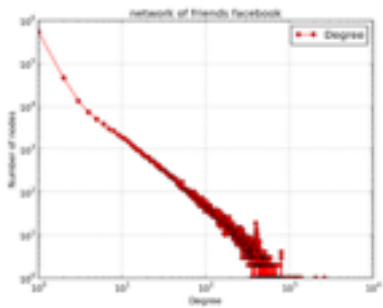

*Figure 1: Degree distribution of the social interaction network*

## 2. Interaction Graph

The Interaction Graph is a directed graph, since it records the presence of an interaction between nodes A and B in a directed way (e.g.: if the interaction is a photo, the network records the fact that A took a photo with B). The number of nodes and edges in the interaction graph is quite low, with respect to the ones of the social graph, as we can see from the summary statistics reported in Table 3. The first empirical observation coming from these data is how the number of interactive friends, inside a social network community, is just a minority with respect to the total number of friends in the network.

*Table 3: Social Interaction Graph Metrics*

| # Nodes | **107518** |
|---|---|
| # Edges | 165501 |
| Average Degree | 1.980 |
| Average Node Degree | 3 |
| Average Clustering Coefficient | 0.0314201008965 |
| Assortativity in-degree | 0.118850333779 |
| Assortativity out-degree | -0.0292121272188 |

Similarly to what we described in Section 2.1, we also built the corresponding Erdos-Renyi random graph, with probability of and edge described as in Equation 1. Resulting clustering coefficient of the random graph obtained was 0.00000307811139486, that is much lower than the clustering coefficient of the interaction graph. Therefore we can conclude that the clustering coefficient of the interaction graph is quite high.

Moreover, being a directed graph, we were able to also obtain the indegree and outdegree assortativity measures. The indegree/outdegree assortativity of a directed network is the tendency of nodes with similar in/outdegree to connect to each other. As we can see from Table 3, in the interaction network the small positive value for the indegree assortativity implies that there is a tendency by the nodes to connect with nodes having similar indegree. On the other hand the negative outdegree assortativity value implies the tendency of nodes to connect with nodes having dissimilar outdegree values. This means that nodes targeted by many connections have the tendency to be connected to nodes that are of the same type. Furthermore we investigated the indegree and outdegree distribution of the interaction graph, and we report in Figure 2 the resulting graph, where we notice a power law distribution also in this case, as reported also for Figure 1.

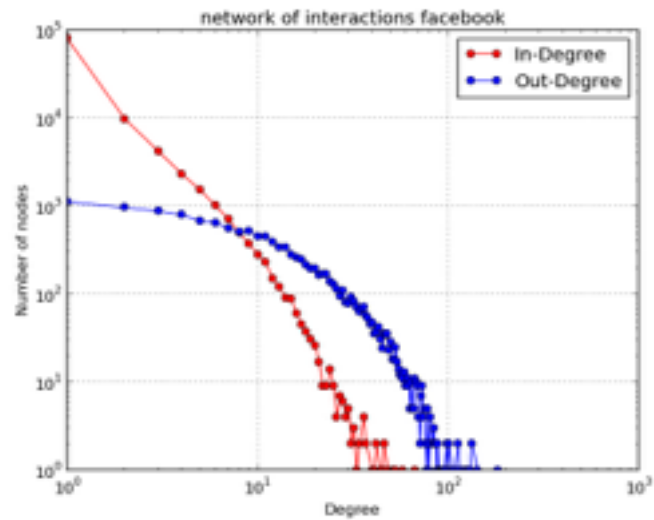

*Figure 2: Indegree and outdegree distribution of the social interaction network*

## 3. Experiments Setup and Results

### 3.1. Comparison between the interaction and the social graph

To make a comparison between the interaction and the social graph, we compared nodes with highest in-degree in the interaction network and nodes with highest degree in the social network. This, in fact, would allow us to understand the possible relationship between the two graphs as well as further elements on the assortativity measures as described in Section 2.1. To perform such comparison we considered the 10 most central nodes in the social graph, the 10 nodes in the social graph with highest degree and the 10 nodes in the interaction graph with highest indegree, and we analyzed the possible correlation between the two sets of nodes. As we can see from Table 4, being one of the top ten nodes in terms of number of friends (one of the ten nodes with highest degree in the social graph), doesn't imply being one of the top ten nodes receiving contacts.

*Table 4: Comparison between the top 10 highest degree nodes in the Social graph and the top 10 highest indegree nodes in the Interaction Graph*

| Id Node Social Graph | Degree Value in the Social Network | Id Node Interact. Graph | Indegree Value in the Interaction Network |
|---|---|---|---|
| **329686** | 2643 | **12130** | 67 |
| **209834** | 2074 | **2259** | 58 |
| **65879** | 1388 | **7445** | 53 |
| **2826** | 1299 | **15015** | 52 |
| **25239** | 1296 | **1148** | 51 |
| **15602** | 1273 | **11892** | 48 |
| **1169** | 1250 | **680** | 46 |
| **2441** | 1207 | **3502** | 46 |
| **483496** | 1200 | **6583** | 45 |
| **15910** | 1157 | **29083** | 44 |

.To further investigate this aspect, we also extracted also the indegree values for the nodes with highest degree in the social graph (the nodes of the 1st column in Table 5), as well as the degree for the highest indegree nodes in the interaction graph (the nodes of 3rd column in Table 5). Results show interesting aspects, and quite unexpected findings. The node 329686 with most connections in the social graph, i.e. with most friends, does not even have a single interaction with other nodes. A possible explanation could be that the node corresponds to a public Facebook page, that has many friends, which however do not interact directly through photos or wall post with anyone.

In any case also the other nodes, in the top ten with the highest degree in the social graph, have very few connections which confirms that many friends do not imply many connections. We also extracted the connected components of the interaction graph showing how, unlike the social graph, the interaction graph is composed by 102963 strongly connected components, creating a very sparse graph matrix.

Table 5: Indegree and degree values for the top ten high scoring nodes in the Social and interaction graph for degree and indegree

| Top ten degree value Nodes Social Graph | Indegree value of the node in the Interaction Graph | Top ten indegree value nodes in the Interaction graph | Degree value of the node in the social graph |
|---|---|---|---|
| **329686** | None | **12130** | 116 |
| **209834** | 1 | **2259** | 226 |
| **65879** | 5 | **7445** | 98 |
| **2826** | 18 | **15015** | 462 |
| **25239** | 10 | **1148** | 98 |
| **25602** | 29 | **11892** | 733 |
| **1169** | 3 | **680** | 159 |
| **2441** | 24 | **3502** | 164 |
| **483496** | 2 | **6583** | 570 |
| **15910** | 4 | **29083** | 101 |

**3.2. Milgram Experiment on the Facebook Dataset**

In order to analyze Small World properties of the Social and Interaction Networks, we decided to replicate the famous Small-World Milgram experiment both on the proposed Social and Interaction Networks [6].

**4.2.1 Milgram Experiment on the social Graph**

To perform Milgram Experiment in the Social Graph, we divided the original set of 657681 nodes in two subsets, which we can set A and set B. Subsequently the source node was randomly chosen from the subset A, and the target node was randomly chosen from the subset B. We decided to implement this procedure to replicate the random procedure used by Milgram for the choice of the source node (the 96 sources were randomly chosen from the telephone directory in the US town of Omaha, in Nebraska) [6]. Moreover also the target we wanted to be chosen randomly, in order to obtain the most likely estimation of the average path length inside the social network graph. Once chosen the source and target nodes, we computed the average paths length between them, and computed the average of all the shortest paths length identified. However, one important aspect must be taken into account. The social graph is a non-connected graph, therefore a path between two randomly selected nodes exists only if they are part of the same connected component. So if two randomly selected nodes are not part of the same connected component, it means that the started chain fails. Therefore, as a further variable, we took into account also the number of failures in the Milgram experiment on the social graph. We performed the experiment on 96, 24000 and 658781 couples of source-target nodes. The number 96 is justified because of the Milgram Experiment, 24000 because of the remake of the Milgram Experiment using email, repeated by Dodds [7] and 658781 because it corresponds to the number of nodes in our social graph. In Table 6 I present the results obtained for the various experiments performed on the Social Graph:

*Table 6: Results of Milgram Experiment on the Social Graph*

| # of couples source-target | Average length of Shortest Path | Percentage of failed chains |
|---|---|---|
| **96** | 5 | 0% (all arrived) |
| **24000** | 5 | 0.02% (5 failures) |
| **657681** | 5 | 0.028% (5 failures) |

The first thing to notice is that the average shortest path is aligned with the six degrees of separation found by the Milgram Experiment. In fact the average path length of 5.9 found by the Milgram Experiment is considered as an upper bound to the geodesic distance (minimum number of edges) between the sources and the targets considered. Therefore, an average path length of 5 perfectly fits within this upper bound definition. Also the remake of Milgram Experiment by Dodds et al in [7], confirmed the average path length of Milgram, and our result of 5, as an average path length, confirms that the social network graph is a small world network. Another interesting point is the number of average failures I obtain. In the Milgram experiment, only 19% reached the intended targets, while in the Dodds' case only 1.5% reached the intended target. In our case the number of failures is computed as the number of non-existing paths between two nodes. A possible justification for this is the way our network is composed. As we showed in Section 1 there is a large component in the graph, that collects almost all the nodes of the entire network. Considering in fact the way the components are made, we can compute the probability of having two nodes coming from two different components. Since draws from of two nodes are independent, the total probability would be:

where:

is the probability of belonging to cluster = #nodes in cluster /tot number of nodes is equal to 0.0028, which means that the probability that two random nodes will belong to two different clusters is approximately 0.28%. This can provide an explanation to the low rate of failures in our Milgram chains.

**4.2.2. Milgram Experiment in the interaction Graph**

We repeated the Milgram experiment described in the previous section also for the interaction graph. Since, as explained in Section 1, the interaction graph is a directed graph we performed the Milgram Experiment in both the original directed graph, and in the undirected version of the interaction graph. As in the previous case, we repeated the experiment for 96, 24000 and 107518 couples (same reasons as before, this time 107518 represents the number of nodes in the graph). Table 7 and Table 8 summarize the results obtained in terms of average path length and percentages of failure in terms of chains.

*Table 7: Results of Milgram Experiment for the directed interaction graph*

| #pairs | Average Path Length | Percentage of Failed chains |
|---|---|---|
| **96** | 8 | 94/96=97% |
| **24000** | 9 | 23562/24000=98% |
| **107518** | 9 | 105723/107518=98% |

*Table 8: Results of Milgram Experiment for the directed interaction graph*

| #pairs | Average Path Length | Percentage of failed chains |
|---|---|---|
| **96** | 7 | 9/96=9.3% |
| **24000** | 7 | 2121/24000=8.8% |
| **107518** | 7 | 9505/107518=8.84% |

The results obtained are extremely promising. First of all what is the interpretation to give to the shortest path between two nodes in the interaction graph? If in the case of the social network, the interpretation was the distance between two people in the network, here the interpretation would be how, following the network of interaction, I can go from one person to another. So in this way the path followed is not through common relationships between the source and the target node, but through common interactions taking place in a path connecting a source node to a target one. So the number of failures increases if the graph is considered as directed since we force the path to be directed and therefore the number of path that could be taken decreases. The average path length increases with respect to the one of the social graph for the same reason. On the other hand, even if the graph is treated as undirected the number of failures is slightly higher than the social case. This is due to the structure of the interaction network that is no longer concentrated around a big component, but sparse in many smaller ones (the number of weakly directed components of the graph in this case is 551004).

**5. Conclusions**

In this paper we analysed a Facebook regional dataset. The dataset is made of two graphs, one describing the friendship among Facebook users and another describing the interactions among the users.
We analysed the two graphs separately, considering network metrics and drawing parallels between the two. Moreover we propose a novel experimental setup to reproduce the Milgram experiment on the two networks, discussing and analyzing deeply the results obtained.
Results are novel and promising, in line with the six degrees of separation. Future work can be implemented, including novel and different social networks to explore their nature and characteristics.